\documentclass[twoside]{elsart}
\usepackage{epsfig}
\usepackage{array}
\usepackage{amssymb,amsbsy,amsmath}
\usepackage{times}
\newcommand{\figref}[1]{Fig.~\protect\ref{#1}}

\renewcommand{\eqref}[1]{Eq.~(\protect\ref{#1})}
\newcommand{\xref}[1]{\protect\ref{#1}}
\newcommand{\fmref}[1]{(\protect\ref{#1})}
\newcommand{\element}[2]{$^{#1}$#2}
\def\V0{\stackrel{\circ}{V}}
\def\v0{\stackrel{\circ}{v}}
\def\half{{\frac{1}{2}}}

\newcommand{\dint}{\mbox{d}}
\renewcommand{\Re}[1]{\mbox{Re}\left(#1\right)}

\newcommand{\tr}{\mbox{tr}}
\newcommand{\op}[1]{%
    \fontdimen12\textfont3=2pt\fontdimen12\scriptfont3=1.4pt%
    \!\null\mathop{\vphantom{#1}\smash{#1}}\limits_{\sim}\null\!}

\newcommand{\Mean}[1]{\big\langle\big\langle \; {#1}\; 
            \big\rangle\big\rangle}

\def\bra#1{\langle \, {#1} \, | \;}
\def\ket#1{\; | \, {#1} \, \rangle}

\def\ketSM{\ket{S\,M}}
\def\braSM{\bra{S\,M}}
\def\ketSMP{\ket{S^\prime\,M^\prime}}

\def\ketsm{\ket{s\,m}}
\def\brasm{\bra{s\,m}}

\def\ketsm1{\ket{s_1\,m_1}}
\def\brasm1{\bra{s_1\,m_1}}
\def\ketsm2{\ket{s_2\,m_2}}
\def\brasm2{\bra{s_2\,m_2}}
\newcommand{\vek}[1]{{\!\vec{\,#1}}}

\begin{document}
%
\typeout{   --- >>>   dimer paper   <<<   ---   }
\typeout{   --- >>>   dimer paper   <<<   ---   }
\typeout{   --- >>>   dimer paper   <<<   ---   }
%
%
\journal{PHYSICA A}
\begin{frontmatter}
\title{Spin dynamics of quantum and classical Heisenberg dimers}
 
\author{D. Mentrup, J. Schnack\thanksref{JS}}
\address{Universit\"at Osnabr\"uck, Fachbereich Physik \\ 
         Barbarastr. 7, 49069 Osnabr\"uck, Germany}
\author{Marshall Luban}
\address{Ames Laboratory \& Department of Physics and Astronomy,
Iowa State University\\ Ames, Iowa 50011, USA}
\thanks[JS]{corresponding author: jschnack\char'100uos.de,\\ http://www.physik.uni-osnabrueck.de/makrosysteme/}

\begin{abstract}
\noindent
Analytical solutions for the time-dependent autocorrelation
function of the classical and quantum mechanical spin dimer with
arbitrary spin are presented and compared.  For large spin
quantum numbers or high temperature the classical and the
quantum dimer become more and more similar, yet with the major
difference that the quantum autocorrelation
function is periodic in time whereas the classical is not.

\vspace{1ex}

\vspace{1ex}

\end{abstract}
\end{frontmatter}
\raggedbottom
\section{Introduction and summary}

There is a growing interest in the magnetic properties of
synthesized molecules \cite{TDP94,GCR94,FST96,LGB97} containing
relatively small numbers of paramagnetic ions. With the ability
to control the placement of magnetic moments of diverse species
within stable molecular structures, one can test basic theories
of magnetism and even begin to explore the design of novel
systems that offer the prospect of useful applications. Most
species of organic-based molecular magnets exhibit very weak
intermolecular magnetic interactions, so that measurements
performed on a bulk sample actually reflect intramolecular
interactions only. The magnetic interaction appears to be well
described by the Heisenberg model with isotropic,
nearest-neighbor exchange. A key quantity is the time- and
temperature-dependent correlation function for pairs of magnetic
moments, as it serves as the basic ingredient for understanding
diverse dynamical phenomena, such as inelastic neutron
scattering \cite{BaL89} and spin lattice relaxation
\cite{Mor56}.

The present study is motivated by a desire to achieve a deeper
understanding of spin dynamics in the Heisenberg model,
especially concerning the trends that occur in arrays of $N$
interacting moments (individual spins $s$) for increasing values
of both $N$ and $s$. The classical Heisenberg model turns out to
provide accurate quantitative results for static properties,
such as magnetic susceptibility, down to thermal energies of the
order of the exchange coupling \cite{Sch99,LLB98}. It is quite
easy to establish the connection of that model, for static
properties, to the corresponding quantum model for arbitrary
$s$. However, considerable care is required to successfully link
up with classical Heisenberg spin dynamics starting from quantum
Heisenberg spin dynamics.

In this article we present the analytical form of the
time-dependent equilibrium auto\-correlation function of the
quantum mechanical dimer with general spin $s$. The trends for
increasing $s$ are explored in some detail and in particular we
compare the quantum results with the exact analytical result
recently derived
\cite{LBC,LuL99} for a classical Heisenberg spin dimer. The
quantum results for arbitrary $s$ are obtained using
Mathematica$^{\circledR}$ to evaluate the Clebsch-Gordan
coefficients.  Results have previously been obtained for
$s=\half$ spin rings of length up to $N=16$ by complete
diagonalization methods \cite{FaL97}, and in Ref.~\cite{MTP81}
some aspects of the high spin limit were discussed.

The present study of the equilibrium autocorrelation function
for large values of $s$ is timely given the fact that NMR
measurements have very recently been performed \cite{LTA99} on a
dimer molecular magnet composed of Fe$^{3+}$ $(s=5/2)$
ions. Heretofore only $s=\half$ dimers have been available for NMR
studies \cite{Kaw66,SvT71,FIK96}. For comparison between theory and
experiment it will also be necessary to incorporate molecular
and single-ion anisotropy terms in the Hamilton operator.  This
will be the subject of a forthcoming article.

\section{The quantum dimer}

The quantum dimer is specified by the Hamilton operator
\begin{eqnarray}
\label{E-2-1}
\op{H}
&=&
\frac{J}{\hbar^2}\;
\op{\vek{s}}_1 \cdot \op{\vek{s}}_2
=
\frac{J}{2 \hbar^2}\;
\left(
\op{\vek{S}}^2 - \op{\vek{s}}_1^2 - \op{\vek{s}}_2^2
\right)
\quad\ ;\quad
\op{\vek{S}}=\op{\vek{s}}_1 + \op{\vek{s}}_2
\ ,
\end{eqnarray}
where $J>0$ describes antiferromagnetic and $J<0$ ferromagnetic
coupling. Throughout this article it is assumed that the spin
quantum numbers of both sites of the dimer are identical,
$s_1=s_2=s$.  The eigenstates $\ketSM$ of total spin
$\op{\vek{S}}^2$
\begin{eqnarray}
\label{E-2-2}
\op{\vek{S}}^2 \ketSM &=& \hbar^2 S (S+1) \ketSM
\ ,\qquad
\op{{S}}_z \ketSM = \hbar M \ketSM
\end{eqnarray}
are also eigenstates of the Hamilton operator with eigenvalues
$E_S$, which, in the absence of a magnetic field, do not depend on
the total magnetic quantum number $M$
\begin{eqnarray}
\label{E-2-3}
\op{H} \ketSM
&=&
\frac{J}{2}\;
\left(
S (S+1) - 2\,s (s+1)
\right) \ketSM
=
E_S\,\ketSM
\ .
\end{eqnarray}
Thus the partition function in the canonical ensemble reads
\begin{eqnarray}
\label{E-2-4}
Z
&=&
\tr\left\{ e^{-\beta\op{H}}\right\} 
=
\sum_{S,M} \braSM e^{-\beta\op{H}} \ketSM
\\
&=&
e^{\beta\,J\,s (s+1)}
\;\sum_{S=0}^{2\,s} (2S+1)\;e^{-\frac{\beta\,J}{2}\,S\,(S+1)}
\nonumber
\end{eqnarray}
and, considering that the Hamilton operator \fmref{E-2-1} is
isotropic, one obtains for the unnormalized autocorrelation
function
\begin{eqnarray}
\label{E-2-5}
\Mean{\op{\vek{s}}_1(t) \cdot \op{\vek{s}}_1(0)}
&=&
\frac{3}{Z}
\sum_{S,M} \braSM \op{{s}}_{1z}(t) \cdot
\op{{s}}_{1z}(0)\, e^{-\beta\op{H}} 
\ketSM
\\
&=&
\frac{3}{Z}
\sum_{S,M,S^\prime,M^\prime} 
e^{\frac{i\,t}{\hbar}(E_S-E_{S^\prime})}\;
e^{-\beta\,E_S}\;
|\braSM\op{{s}}_{1z}\ketSMP|^2
\nonumber
\ .
\end{eqnarray}
\begin{figure}[t]
\begin{center}
\epsfig{file=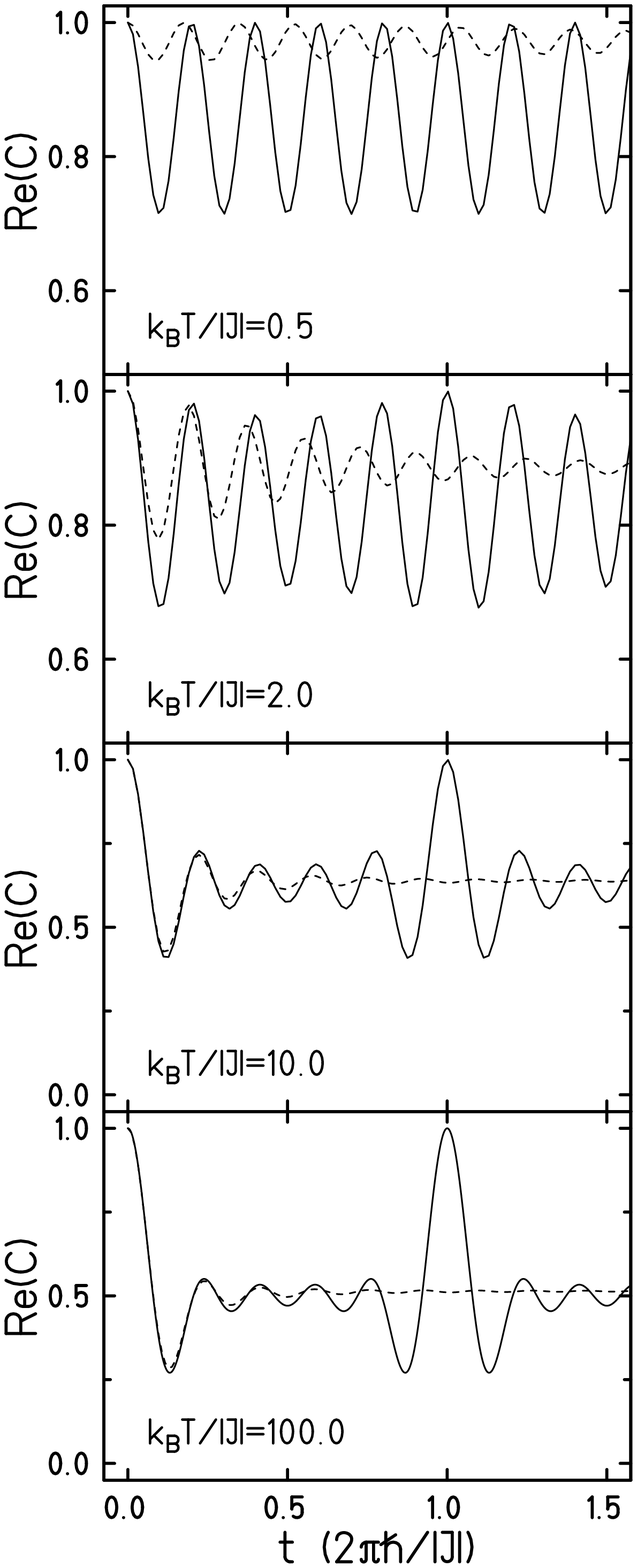,height=150mm}
$\qquad$
\epsfig{file=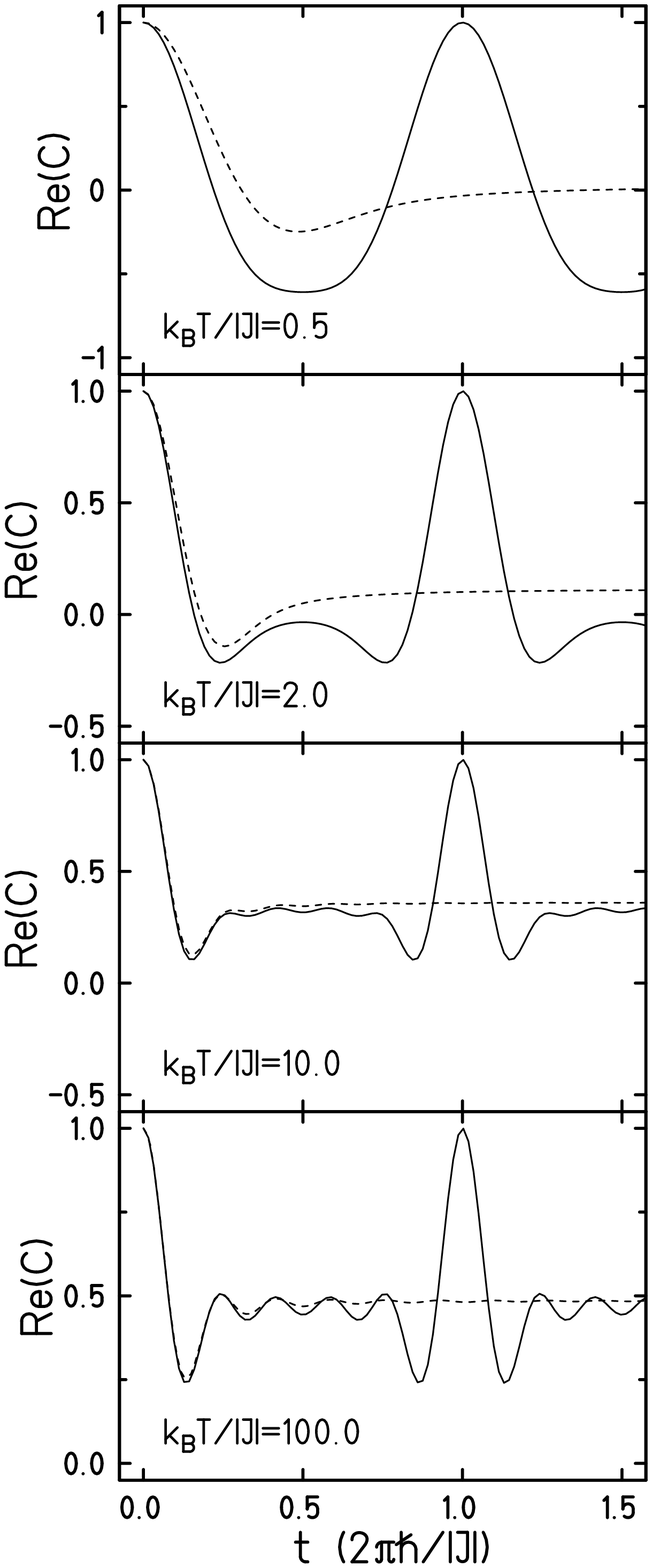,height=150mm}
\caption{Normalized autocorrelation function $\Re{C(t)}$, 
see \eqref{E-A-2}, for a spin-$\frac{5}{2}$-dimer for four
different temperatures (solid lines). The left panels display our
results for the ferromagnetic dimer, the right panels the
antiferromagnetic case \cite{MeS99}. The dashed lines show the
classical result; for details see the next section.}
\label{F-2-1}
\end{center} 
\end{figure} 

The last expressions simplifies when we take into account that only
matrix elements with $M=M^\prime$ and a difference in total spin
not larger than one contribute \cite{CoS64}, i.e.
\begin{eqnarray}
\label{E-2-6}
\braSM\op{{s}}_{1z}\ketSMP = 0
\qquad\mbox{if}\qquad
|S-S^\prime| > 1
\qquad\mbox{or}\qquad
M\ne M^\prime
\ .
\end{eqnarray}
The resulting expression is shown in \eqref{E-A-1} in the
appendix.  Note that the angular frequency spectrum of the spin
dimer is given by integer multiples of ${J}/{\hbar}$, which in
turn means that the autocorrelation function is periodic in time
and that the recurrence time $\tau$ only depends on the
coupling $J$ but not on the spin quantum number $s$
\begin{eqnarray}
\label{E-2-8}
\omega \in \left\{ \frac{J}{\hbar} S \right\}
\ , \quad S=0,\dots , 2\,s
\quad\Rightarrow\quad
\tau = \frac{2\,\pi\,\hbar}{J}
\ .
\end{eqnarray}
This is of course true for all Hamilton operators that can be
written like the term on the r.h.s. of \eqref{E-2-1}, namely for
the spin trimer and the spin tetrahedron.

Figure \xref{F-2-1} shows the autocorrelation function
normalized to unity at $t=0$ for a spin-$\frac{5}{2}$-dimer, a
system that has been synthesised (Fe dimer) and that is
currently under investigation \cite{LTA99}.  The analytical
expression for this autocorrelation function is given in
\eqref{E-A-2} in the appendix. One clearly sees that the
autocorrelation function, which is a superposition of five
harmonic oscillations and a constant, is dominated at low
temperatures by the highest frequency in the ferromagnetic case
and by the lowest frequency in the antiferromagnetic case. At
higher temperatures other frequencies also contribute. One also
notices that, independent of temperature, the autocorrelation
function returns to its initial value after $\tau =
\frac{2\,\pi\,\hbar}{J}$.

\section{Comparison to the classical dimer}

In order to compare the results of the quantum dimer for
different spin quantum numbers $s$ with each other and with the
classical dimer it is useful to introduce normalized spin operators
\begin{eqnarray}
\label{E-3-0}
\op{\vek{\epsilon}}_n
=
\frac{\op{\vek{s}}_n}{\sqrt{\hbar^2\,s(s+1)}}
\ ,\quad
n=1,2
\quad \ ,
\end{eqnarray}
which depend on $s$. Note, that 
\begin{eqnarray}
\label{E-3-10}
\left[
\op{\epsilon}_{n\,x} ,\op{\epsilon}_{n\,y} 
\right]
=
\frac{i}{\sqrt{s(s+1)}}\;
\op{\epsilon}_{n\,z}
\ ,
\end{eqnarray}
and hence these become commuting operators for
$s\rightarrow\infty$. 
\eqref{E-3-0} suggests that we define a classical Hamilton
function $H_{c}$
\begin{eqnarray}
\label{E-3-1}
H_{c}
=
J_{c}\;\vek{e}_1 \cdot \vek{e}_2
\quad , \quad
J_{c} = J\,s(s+1)
\ ,
\end{eqnarray}
where $\vek{e}_1$ and $\vek{e}_2$ are unit vectors
(c-numbers). We expect that the thermal properties of this
classical Heisenberg system will coincide with those of the
quantum Heisenberg dimer if $s\gg 1$ except for very low
temperatures. This is because the spectrum of eigenvalues of
$\op{\epsilon}_{n\,z}$ is confined within $(-1,1)$ and becomes
dense for $s\rightarrow\infty$ and thus coincides with the
continuous range of ${e}_{n\,z}$.

Similarly, if we substitute \eqref{E-3-0} in the quantum
equations of motion for $\op{\vek{s}}_1$ and $\op{\vek{s}}_2$,
we have
\begin{eqnarray}
\label{E-3-2}
\dot{\op{\vek{\epsilon}}}_1
=
-\Omega\;
\op{\vek{\epsilon}}_1 \times \op{\vek{\epsilon}}_2
\quad , \quad
\dot{\op{\vek{\epsilon}}}_2
=
+\Omega\;
\op{\vek{\epsilon}}_1 \times \op{\vek{\epsilon}}_2
\end{eqnarray}
where
\begin{eqnarray}
\label{E-3-3}
\Omega
=
\frac{J\,\sqrt{s(s+1)}}{\hbar}
=
\frac{J_{c}}{\hbar\,\sqrt{s(s+1)}}
\ .
\end{eqnarray}
This suggests that we prescribe the following equations of
motion for the classical unit vectors $\vek{e}_1$ and
$\vek{e}_2$, 
\begin{eqnarray}
\label{E-3-11}
\dot{\vek{e}}_1
=
-\Omega\;
\vek{e}_1\times\vek{e}_2
\quad , \quad
\dot{\vek{e}}_2
=
+\Omega\;
\vek{e}_1\times\vek{e}_2
\ .
\end{eqnarray}
We emphasize that in these equations $\Omega$ is given by
\eqref{E-3-3}. It is expected that the autocorrelation function
derived using \fmref{E-3-11} and the canonical ensemble average
based on $H_c$ will coincide in the large $s$ limit with the
normalized autocorrelation derived from \eqref{E-A-1}. This
expectation is in fact confirmed as discussed below.

Using the fact that the total spin is a constant of motion the
classical partition function can be derived as
\cite{Fis64,CLA99}
\begin{eqnarray}
\label{E-3-4}
Z_{c}
&=& 
\half \int_0^2 dS \, S \,
\exp\left\{-\frac{\beta\,J_{c}}{2} \, (S^2-2)\right\}
\\
&=&
\frac{1}{2\,J_{c}}\;\int_{-J_{c}}^{J_{c}}\,\dint E\;
\exp\left\{-\beta\,E \right\}
=
\frac{\sinh(\beta\,J_{c})}{\beta\,J_{c}}
\nonumber
\ .
\end{eqnarray}
Note that the classical density of states turns out to be a
constant in the energy interval $[-J_{c},J_{c}]$.  This
coincides nicely with the quantum density of states which can be
obtained by counting the discrete eigenvalues per unit energy
interval and normalizing the density so that its integral gives
$1$.
One can show, starting from \eqref{E-2-4}, that the quantity
$Z/(4\,s(s+1))$ is in close numerical agreement with $Z_c$ for
temperatures $k_B\,T\,>\,0.2\,J\,s(s+1)$. This serves to
clearly define the classical regime for the thermal properties
of the dimer.

\begin{figure}[t]
\begin{center}
\epsfig{file=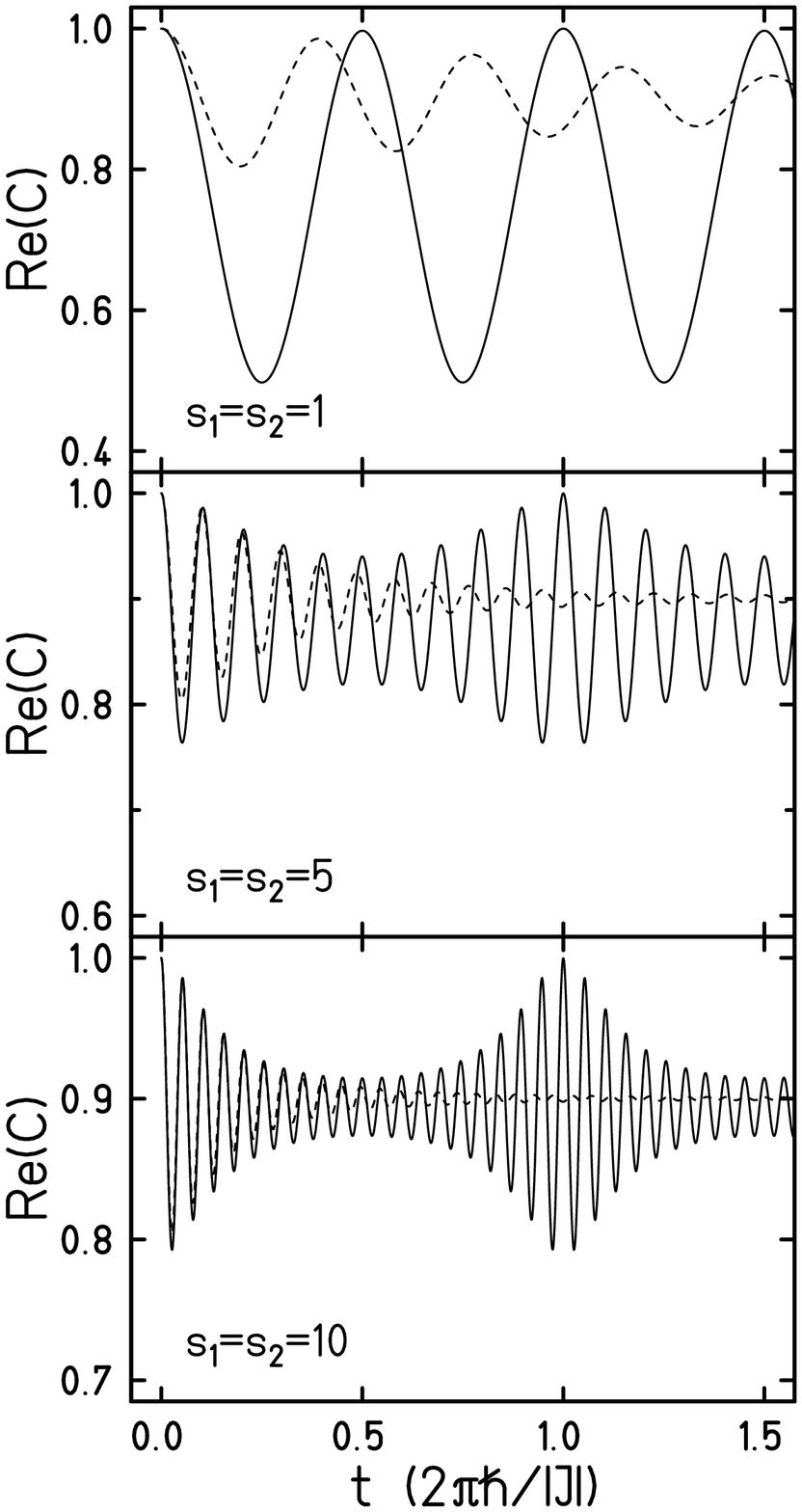,height=120mm}
$\qquad$
\epsfig{file=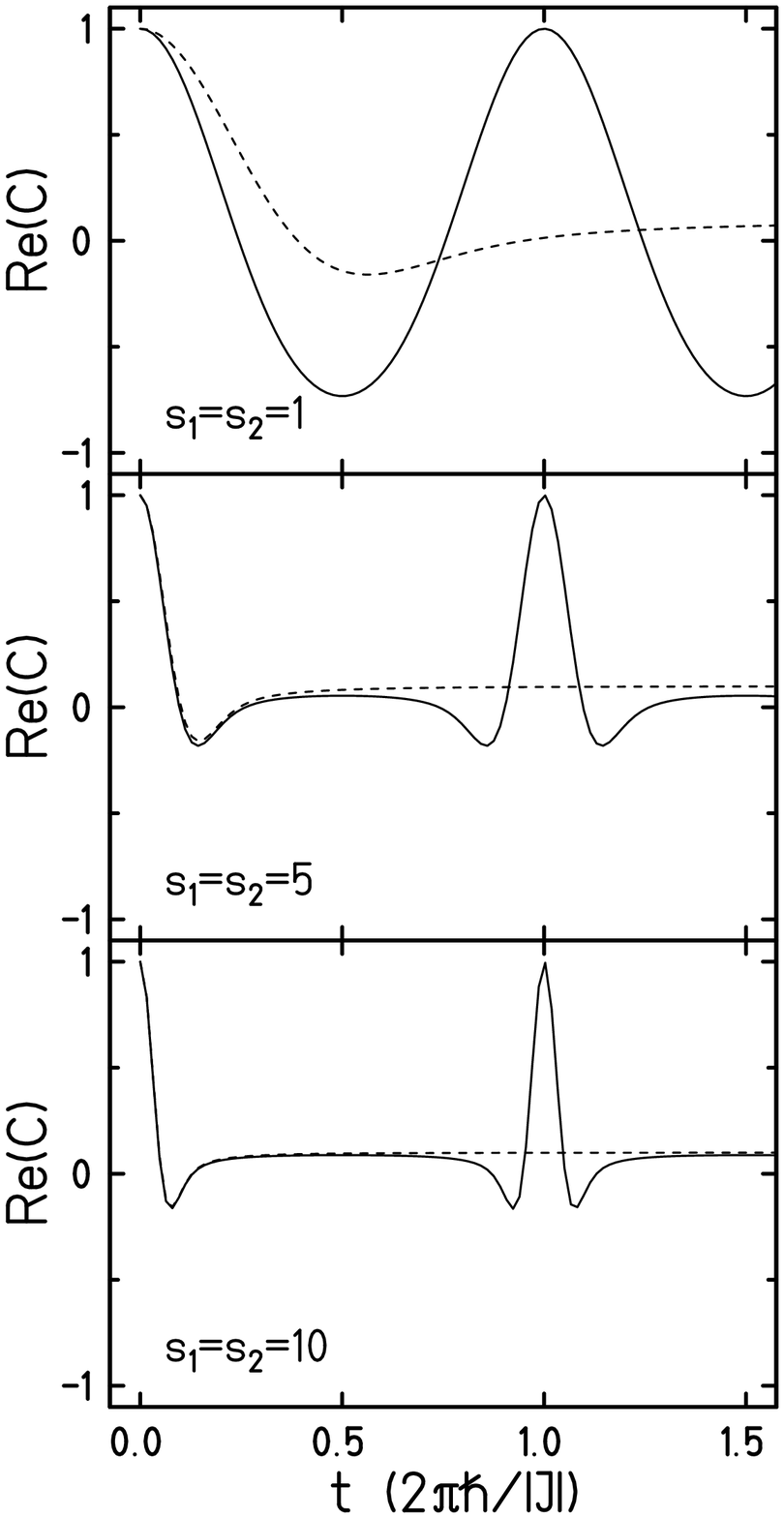,height=120mm}
\caption{Normalized autocorrelation function for three different spins at
the temperature $k_B\,T/(J\,s\,(s+1))=0.2$. The left panels displays the
ferromagnetic dimer, the right panels the antiferromagnetic one. The
solid lines show the quantum result, the dashed lines the classical. }
\label{F-3-1}
\end{center} 
\end{figure} 

For the classical autocorrelation function one finds that
\cite{LBC,LuL99} 
\begin{eqnarray}
\label{E-3-5}
C_{c}(t)
&=&
\half
\left[
1-\coth(\beta\,J_{c})+\frac{1}{\beta\,J_{c}}
\right]
+
\frac{\beta\,J_{c}}{1-\exp(-2\,\beta\,J_{c})}
\\
&&
\times\int_0^2\dint S\;
S\left(1-\frac{S^2}{4}\right)
\exp\left(-\frac{\beta\,J_{c}\,S^2}{2} \right)
\cos\left(S\,\Omega\,t\right)
\nonumber
\ ,
\end{eqnarray}
which can be integrated using error functions of complex
arguments, see Ref.~\cite{MeS99}.  In contrast to the quantum
autocorrelation function \fmref{E-2-5}, the classical quantity
is real. The reason is that $\op{\vek{s}}_1(t) \cdot
\op{\vek{s}}_1(0)$ is not a hermitian operator for $t\ne 0$. If
one would like to construct a hermitian operator,
$\half\left(\op{\vek{s}}_1(t) \cdot \op{\vek{s}}_1(0)
+\op{\vek{s}}_1(0) \cdot \op{\vek{s}}_1(t)\right)$ would be
appropriate. This coincides with the real part of our definition
Eqs.~\fmref{E-2-5} and \fmref{E-A-1}. It is also interesting to
note that the imaginary part of $\Mean{\op{\vek{s}}_1(t) \cdot
\op{\vek{s}}_1(0)}$ does indeed vanish in the high temperature
limit.

In \figref{F-3-1} the dashed curves display the classical
autocorrelation function obtained from
\fmref{E-3-5} together with the quantum result
(solid lines) for three different spin quantum numbers.  In
order to compare the different correlation functions all
spectra have been mapped on the same energy interval
$[-J_c,J_c]$. Thus the different figures show the
autocorrelation functions for the same position of the mean
excitation energy in the spectrum, i.e. the same
\begin{eqnarray}
\frac{k_B\,T}{J\,s\,(s+1)}
=
\frac{k_B\,T}{J_c}
=
0.2
\ .
\end{eqnarray}
Based on our earlier remark concerning the close numerical
agreement of the classical and quantum partition functions when
$k_B\,T\,>\,0.2\,J_c$, we anticipate similar agreement for the
autocorrelation function in this temperature range. This is
confirmed on inspecting the various panels of \figref{F-3-1},
which demonstrate nicely that the quantum autocorrelation
function approaches the classical result with increasing $s$.
The most prominent difference between these results is that the
classical autocorrelation function does not return to its
initial value but approaches a unique non-zero limit, whereas
the quantum autocorrelation function is recurrent with a
recurrence time independent of spin and temperature.  This is
due to the fact that the classical system has a continuous
spectrum of excitations in the angular frequency interval 
$[0, 2\Omega]$ whereas the quantum system possesses a discrete
spectrum of excitations which are all integer multiples of the
lowest one.


{\bf Acknowledgments}\\[5mm] 
The authors would like to thank
K.~B\"arwinkel and H.J.~Schmidt for valuable
discussions. M.L. would like to thank the Department of Physics
of the University of Osnabr\"uck for the warm hospitality
extended to him during his visit when a part of this work was
performed.  The Ames 
Laboratory is operated for the United States Department of
Energy by Iowa State University under Contract
No. W-7405-Eng-82.
\appendix
\section{The quantum dimer}

Using the matrix properties \fmref{E-2-6} the unnormalized
autocorrelation function \fmref{E-2-5} can be simplified
to
\begin{eqnarray}
\label{E-A-1}
&&
\Mean{\op{\vek{s}}_1(t) \cdot \op{\vek{s}}_1(0)}
=
\frac{3}{Z}\;
e^{\beta\,J\,s (s+1)}
\\
&&\times
\Bigg\{
\sum_{S,M} 
e^{-\frac{\beta\,J}{2}\,S (S+1)}
|\braSM\op{{s}}_{1z}\ketSM|^2
\nonumber \\
&&\quad
+
\sum_{S=1}^{2 s} 
\sum_{M=-S+1}^ {M=S-1}
|\braSM\op{{s}}_{1z}\ket{S-1\,M}|^2
\nonumber \\
&&\qquad\times
\Big(
\cos\left[\frac{t\,J}{\hbar} S\right]
\left[
e^{-\frac{\beta\,J}{2} S (S+1)}
+e^{-\frac{\beta\,J}{2}\,S (S-1)}
\right]
\nonumber \\
&&\qquad\quad
+
i \sin\left[\frac{t\,J}{\hbar} S\right]
\left[
e^{-\frac{\beta\,J}{2} S (S+1)}
-e^{-\frac{\beta\,J}{2}\,S (S-1)}
\right]
\Big)
\Bigg\}
\nonumber
\ .
\end{eqnarray}
Taking as an example the case $s=5/2$ yields
\begin{eqnarray}
\label{E-A-2}
&&
C(t) 
=
\frac{\Mean{\op{\vek{s}}_1(t) \cdot \op{\vek{s}}_1(0)}}
     {\Mean{\op{\vek{s}}_1(0) \cdot \op{\vek{s}}_1(0)}}
=
\\&&
\Bigg[
330 
+ 
180\,{e^{{{5\,J\,\beta}}}} 
+ 
84\,{e^{{{9\,J\,\beta}}}} 
+ 
30\,{e^{{{12\,J\,\beta}}}} 
+ 
6\,{e^{{{14\,J\,\beta}}}} 
\nonumber \\
&&
+
35\,{e^{{{J}\,\left( {\frac{i\,t}{\hbar}}
                + {{14}\beta}\right) }}} 
+ 
35\,{e^{{{J}\,\left( {-\frac{i\,t}{\hbar}}
                + {{15}\beta} \right) }}} 
+ 
64\,{e^{{{2\,J}\, \left( \frac{i\,t}{\hbar} + 6 \beta \right) }}} 
+ 
64\,{e^{{{2\,J}\,\left( {-\frac{i\,t}{\hbar}}
                + {{7}\beta}
                \right) }}} 
\nonumber \\
&&
+ 
81\,{e^{{{3\,J}\,\left( \frac{i\,t}{\hbar} + 3 \beta \right) }}} 
+ 
81\,{e^{{{3\,J}\, \left( -\frac{i\,t}{\hbar} + 4 \beta \right) }}} 
+ 
80\,{e^{{{J}\,\left( {\frac{4\,i\,t}{\hbar}}
              + {{5}\beta}\right) }}} 
+ 
80\,{e^{{{J}\,\left( -\frac{4\,i\,t}{\hbar} + 9 \beta \right) }}}
\nonumber \\
&&
+
55 {e^{{\frac{{{5\,i}}\,J\, t}{\hbar}}}}
+ 
55\,{e^{{{5\,J}\, \left( -\frac{i\,t}{\hbar} + \beta
\right) }}} 
\Bigg]
\nonumber\\&&
\Bigg/
\Bigg[
{35\,\left( 11 + 9\,{e^{{{5\,J\,\beta}}}}
+ 7\,{e^{{{9\,J\,\beta}}}} 
+ 5\,{e^{{{12\,J\,\beta}}}} 
+ 3\,{e^{{{14\,J\,\beta}}}}
+ {e^{{{15\,J\,\beta}}}}
        \right) }
\Bigg]
\nonumber
\ .
\end{eqnarray}
Autocorrelation functions for other spin quantum numbers can be
evaluated using a Mathematica$^{\circledR}$ 3.0 script, that the reader is
encouraged to download from our web site \cite{MeS99}.

\end{document}